\documentclass[aps,physrev,preprint,superscriptaddress]{revtex4-2}

\usepackage{graphicx}
\usepackage{subfig}
\usepackage{floatrow}
\usepackage{caption}
\usepackage[colorlinks,citecolor=red,linkcolor=blue]{hyperref}

\begin{document}

\graphicspath{{figures/}}


\title{Topologically trivial semiconducting behavior and polaronic effects in antiferromagnetic $EuZn_2As_2$ and $EuCd_2Sb_2$}


\author{Divyanshi Sar*}
\affiliation{Department of Physics, Applied Physics and Astronomy, Binghamton University, Binghamton, New York 13902, USA}
\author{Mingda Gong*}%
\affiliation{Department of Physics, Applied Physics and Astronomy, Binghamton University, Binghamton, New York 13902, USA}
\author{Tetiana Romanova}%
\affiliation{Institute of Low Temperature and Structure Research, Polish Academy of Sciences, Okólna 2, 50-422 Wrocław, Poland}
\author{Luka Khizanishvili}%
\affiliation{Department of Physics, Applied Physics and Astronomy, Binghamton University, Binghamton, New York 13902, USA}
\author{Hannah Park}%
\affiliation{Department of Physics, Applied Physics and Astronomy, Binghamton University, Binghamton, New York 13902, USA}
\author{Dariusz Kaczorowski}%
\affiliation{Institute of Low Temperature and Structure Research, Polish Academy of Sciences, Okólna 2, 50-422 Wrocław, Poland}
\author{Wei-Cheng Lee}%
\affiliation{Department of Physics, Applied Physics and Astronomy, Binghamton University, Binghamton, New York 13902, USA}
\author{Pegor Aynajian}%
\affiliation{Department of Physics, Applied Physics and Astronomy, Binghamton University, Binghamton, New York 13902, USA}
\begin{abstract}
The Eu-based $EuA_2X_2$ (A = Zn, Cd, In, Sn; X = P, As, Sb) family of compounds has recently attracted significant attention as a promising platform for exploring magnetic topological materials, with several members either predicted or reported to exhibit nontrivial topological properties. We investigate the previously reported topological semimetals, $EuZn_2As_2$ and $EuCd_2Sb_2$, using scanning tunneling microscopy and spectroscopy, complemented by various first-principles computational approaches. Through examination of the cleaved surfaces, step-edges, and defect states, we determine the trivial semiconducting behavior in both material systems, with no evidence of topological surface or edge states. These experimental results are consistent with our theoretical analysis revealing the absence of topological band inversion in either system. Our experimental observations also reveal numerous intrinsic defects that trap charge carriers. These defects may facilitate the formation of magnetic polarons, providing a natural explanation for the colossal negative magnetoresistance observed in many of the $EuA_2X_2$ material systems.
\end{abstract}


\maketitle

Magnetic topological materials represent a fascinating frontier in condensed matter physics, offering a rich array of fundamental phenomena, including the axion insulator~\cite{Li2010,Mogi2017,Liu2020} and the quantum anomalous Hall states~\cite{Liu2008,Chang2015,Deng2020}. These materials not only deepen our understanding of topological phases but also open up unprecedented opportunities for technological advancement~\cite{Nayak2008,AdyStern2013,40yearsofquantumcomputing}.

Over the past decade, several families of intrinsic antiferromagnetic topological material systems have been theoretically predicted. Among them, the $EuA_2X_2$ family (where A represents a metal such as Zn, Cd, In, Sn; and X represents P, As, Sb) has attracted quite some interest~\cite{Rahn2018,HangLi2019,Xu2019,Sato2020,Riberolles2021,Gong2022,Goforth2008,YangWang2022,Taddei2022}. The variation in crystal structure of the different compounds within this family, combined with the tunability of electronic correlation (between 3d and 4d electronic states) and the magnitude of the spin-orbit coupling (between As and Sb) make these material systems an excellent platform to investigate the interplay between correlation, topology, and magnetism~\cite{YangWang2022,Taddei2022,Soh2019,Jo2020,Su2020,Gati2021,Xu2021,Blawat2022,Ohno2022,Blawat2023,Yu2023}. Moreover, the colossal negative magnetoresistance observed in several {$EuA_2X_2$} material systems, makes them particularly suitable for potential spintronic applications\cite{Goforth2008,ZhiChenWang2022,HCChen2020,Luo2023,Pervakov2024,YingLi2024,Balguri2024,Pohlit2018,Calderon2004,Alecrivillero2023,Souza2022,cook2025,kopp2025}.

Theoretical studies and first principles calculations have revealed the topological nature in these material systems, marked by inverted bulk bands protected by the crystalline symmetries. {$EuIn_2As_2$} and {$EuSn_2As_2$} have been predicted to be topological axion insulators~\cite{HangLi2019,Xu2019,lv_chen_zhang_jiang_zhong_2022,Sarkar2022}, whereas the isostructural $EuCd_2As_2$ , $EuZn_2As_2$ and $EuCd_2Sb_2$ have been proposed to be magnetic Weyl semimetals~\cite{Rahn2018,YangWang2022,Taddei2022,Soh2019,Su2020,Blawat2023,Yu2023,Yi2023}. Early experimental investigations, including transport measurements, scanning tunneling microscopy and spectroscopy (STM), and angle resolved photoemission spectroscopy (ARPES) in these systems have provided some evidence supporting the predicted axion insulator~\cite{Sato2020,Riberolles2021,Gong2022,HCChen2020,Zhang2020,Yu2020,Regmi2020,Li2021,Zhao2021,Shinozaki2021,Yan2022} and magnetic Weyl semimetal state~\cite{Taddei2022,Jo2020,Su2020,Ohno2022,Blawat2023,Yi2023,Cao2022,Ma2019}.

Recent studies have challenged the classification of $EuCd_2As_2$ as a magnetic topological material. Compelling experimental evidence has unveiled its topologically trivial semiconducting nature~\cite{Nishihaya2024,Nelson2024,Santos-Cottin2023,Nasrallah2024}. This finding prompts an important question: do the isostructural sibling compounds, $EuCd_2Sb_2$ and $EuZn_2As_2$, also exhibit topologically trivial antiferromagnetic semiconducting behavior?

Here, using STM complemented with first principles calculations, we study the low energy electronic states in $EuCd_2Sb_2$ and $EuZn_2As_2$. Despite previous theoretical and experimental evidence of non-trivial topology~\cite{Su2020,Blawat2023,Luo2023,Yi2023}, our experiments and first principles calculations reveal the trivial semiconducting behavior in both compounds in analogy with the findings in $EuCd_2As_2$~\cite{Nishihaya2024,Nelson2024,Santos-Cottin2023}. Furthermore, by comparing various first-principles calculation approaches (DFT+U, SCAN+U, mBJ+U), we provide a comprehensive understanding of the role of electron correlations in these material systems. This analysis not only explains our experimental observations but also clarifies inconsistencies observed in previous theoretical studies. All together, these findings provide direct evidence of the trivial magnetic semiconducting behavior in isostructural $EuZn_2As_2$ and $EuCd_2Sb_2$. Finally, our experimental topographs and conductance maps reveal a key insight into the unusual electronic behavior in these compounds. Through examination of the intrinsic defects in $EuZn_2As_2$ , we discover that they locally trap charge carriers within a few nanometers of the defect sites. In a magnetic background, these localized charges can polarize surrounding magnetic moments forming magnetic polarons. This provides a natural explanation of the source of the large negative magnetoresistance observed in many of these compounds~\cite{Goforth2008,ZhiChenWang2022,HCChen2020,Luo2023,Pervakov2024,YingLi2024,Balguri2024,Pohlit2018,Calderon2004,Alecrivillero2023,Souza2022,cook2025,kopp2025}.
\vspace{0.1cm}

\section*{Results and Discussion}
\label{sec:results&discussion}

Single crystals of $EuZn_2As_2$ and $EuCd_2Sb_2$ were grown from Sn flux. Details of the growth and characterization are described in the Supplemental Material. Figures 1a,1b show the P-3m1 crystal structure of $EuA_2X_2$ (A=Cd, Zn; X=Sb, As). It has in-plane hexagonal structure with alternating Eu and {$A_2X_2$} layers. The $Eu^{2+}$ ions carry the 4f magnetic moment and are mostly localized in this family of materials, driven by the strong electronic correlations. Prior to the STM experiments, the crystals are cleaved in situ in ultra-high vacuum at room temperature and immediately transferred to the microscope head, which is held at a temperature of  $\sim$9K. Figure 1c shows a large scale topographic image of the exposed surfaces of $EuZn_2As_2$. The strong covalent bonding within the As-Zn bilayers is difficult to break, thus making the Eu-As bond energetically the most likely cleaving plane (shown in Figure 1a). The cleaving between Eu-As layers was also observed in our previous work on $EuIn_2As_2$~\cite{Gong2022}. The cleaving thus exposes either the Eu- or the As-terminated surface. In our experiments, we observe a single type of surface with atomic steps that are multiples of the c-axis unit cell. The structural similarity between Eu- and As-terminations makes surface identification challenging. However, by comparing the $EuZn_2As_2$ topographs and defect patterns with our experimental topographic data of $EuCd_2Sb_2$ (Figure 3a), we can reasonably infer that the topographs in Figures 1c,1d correspond to the As-terminated surface, as discussed below.

The topographic image displayed in Figure 1c shows a large number of triangular defects (see Figure 1d). These blurry triangular defects, as observed by the STM tip, span a lateral size of several nanometers. In addition to these rather profound defects, a few other random types of defects, which occur much less frequently, are also observed (see Figure 1d). By focusing initially on regions free of defects, the STM differential conductance (dI/dV) spectrum, which reflects the local electronic density of states, reveals a relatively large energy gap of approximately 1.2 eV with a flat bottom (Figure 1e). This observation suggests that {$EuZn_2As_2$} behaves as a semiconductor. By comparing the spectra measured above and below {$T_N$} = 19K (Figure 1e), we identify that magnetism does not play a significant role in the observed semi-conducting gap behavior. The flat bottom of the gap with zero dI/dV conductance indicates that no in-gap surface states are observed at either temperature. Similar dI/dV measurements obtained along a line crossing a single unit-cell step edge, as shown in Figure 1f, reveals a relatively uniform gap structure across the step with no observable surface or edge states. A similar large bandgap was also recently seen at 77K~\cite{Kong2024}.

We now turn our focus to examining the influence of the observed defects on the low-energy electronic states. Figures 2a,2b show the two relatively rare defects (the three-lobe protrusion (Figure 2a) and the triangular depression (Figure 2b)) to have negligible effect on the electronic states near the Fermi energy. The STM dI/dV spectra measured across each defect through its center show a rather uniform conductance (see Supplementary Fig.S2 for additional spectroscopic-imaging maps). On the contrary, the more frequently observed large triangular defects show a rather rich in-gap state behavior near the Fermi energy. Spectra taken near the defect reveal multiple peaks at different energies within the semiconducting gap (Figure 2c). The peak energies have a strong dependence on the position of the tip with respect to the defect, as seen in the spectra taken across the defect (orange line in inset of Figure 2c). Spectroscopic imaging on a 5x5 $nm^2$ area over the defect shows the in-gap states to form energy-dependent sharp bow- or ring-like structures with enhanced differential conductance around the defect (insets in Figure 2c).

Ring-like structures are commonly observed in STM experiments on doped 
semiconductors and insulators~\cite{Brar2011,Lee2011,DiBernardo_2022,LeQuang_2018,Morgenstern2001}. These structures arise from the insufficient electronic screening of the tip-induced electric field, which allows it to penetrate the material’s surface. The localized electric field resulting from the applied bias potential difference (as well as the difference between the work functions of the tip and sample) modifies the charge distribution at the semiconductor’s surface, causing the electronic bands (valence and conduction) to bend. The phenomenon is known as tip-induced band bending (TIBB)~\cite{Brar2011,Lee2011,DiBernardo_2022,LeQuang_2018,Morgenstern2001}. Consequently, dispersing ring-like features arise near defects due to the ionization of the defect states induced by the TIBB~\cite{Brar2011,Lee2011,DiBernardo_2022,LeQuang_2018,Morgenstern2001}, indicating trapped charge carrier at the defect sites. Figure 2d displays STM conductance maps on an area of 5x5 $nm^2$ at various energies showing the dispersion of the ring-like structures with applied bias-voltage. Some of the dispersing features are induced by nearby defects located near or outside the perimeter of the field of view (see supplementary movie). The observed ring structures induced by the ionization of the defect states (Figure 2d) and the flat-bottomed gap structure seen on the surfaces and atomic step-edges (Figure 1f), altogether indicate the absence of any sort of conductive surface states and demonstrate the topologically trivial, semiconducting nature of $EuZn_2As_2$.

{$Eu^{2+}$} vacancies is the most energetically stable point-defect in all {$EuA_2X_2$} compounds~\cite{Jo2020,Nelson2024,Zhang2019}, and are responsible for the persistent hole-doping behavior observed across this material family~\cite{Sato2020,Soh2019,Jo2020,Cao2022,Nelson2024}. The effects of the hole doping is evident in the tunneling spectra shown in Figures 1e,1f, where the Fermi level is shifted downwards toward the edge of the valence band. This shift makes the material appear semi-metallic in the resistivity measurements (see supplementary Fig.S1 and~\cite{Blawat2022}. See also~\cite{Santos-Cottin2023} for the case of $EuCd_2As_2$).  Our direct visualization of trapped charge carriers at these intrinsic defect sites offers valuable information about the magnetotransport properties of this material, which is another common feature among many of these compounds~\cite{Goforth2008,ZhiChenWang2022,HCChen2020,Luo2023,Pervakov2024,YingLi2024,Balguri2024,Pohlit2018,Calderon2004,Alecrivillero2023,Souza2022,cook2025,kopp2025}. In a magnetic system, the spin of a trapped charge (defect) can polarize the surrounding magnetic moments, forming ferromagnetic clusters commonly referred to as magnetic polarons~\cite{Durst2002,YongJiang_2009}. Eu-based magnetic semiconductors are particularly susceptible to magnetic polaron formation due to their strong exchange coupling and low carrier density~\cite{Pohlit2018,Calderon2004,Shon2019,Torrance1972}. Recent magnetotransport measurements provide evidence for the formation of such clusters~\cite{Blawat2022,Blawat2023,YingLi2024,Zhang2020}. Furthermore, another study has shown that increasing the amount of Eu vacancies to the several percent level transforms the AFM ground state into a long range ferromagnetic order~\cite{Jo2020}. Taken together, these observations point to the formation of magnetic polarons, arising from trapped charge carriers at defect sites~\cite{Alecrivillero2023,Souza2022,cook2025,kopp2025}. We emphasize that the rings observed in our STM conductance maps do not represent a direct image of a polaron, but rather the signature of a trapped charge carrier.
          
Next, we turn our attention to investigating the sibling compound {$EuCd_2Sb_2$}, which is isostructural and isoelectronic to {$EuZn_2As_2$}. Figures 3a,3b show topographic images of the exposed flat surface, sharing direct resemblance with the surface seen in Figures 1c,1d of {$EuZn_2As_2$}. While we frequently see single c-axis unit cell atomic steps, occasionally, we also find areas with sub-unit cell atomic steps (Figure 3c). From the observed sub-unit cell heights and their asymmetry, we can identify the chemical terminations, as depicted in Figure 3d. Specifically, a $\approx 2Å$ thick step-height can only correspond to an Eu-terminated surface by breaking the same bonds with the layers above and below, whereas a $\approx 5Å$ thick step-height indicates a Cd-Sb bilayer.

STM differential conductance (dI/dV) spectra taken on each surface away from defects, shown in Figure 3e, reveal a semiconducting like behavior with gapped electronic states near the Fermi energy. The magnitude of the observed gap is smaller than that seen on {$EuZn_2As_2$} (Figure 1f). Below the Fermi energy, the spectra show finite density of states (conductance), particularly on the Sb-Cd surface. A dramatic increase of the density of states are seen on both surfaces for energies nearing -1 eV, which corresponds to the onset of the f-electronic density of states (further discussed below). The contrast between the density of states on both surfaces can also be seen by taking the spectra (Figure 3f) along a line crossing the two surfaces (arrow in Figure 3c). Overall, on both the surfaces and across each sub-unit cell step, the spectra show the corresponding gapped structure with no observable edge states. 
     
The impact of the triangular defects on the electronic density of states in {$EuCd_2Sb_2$} is significantly weaker as compared to those seen in {$EuZn_2As_2$} (see Figure 2 and Supplementary Figs. S3-S7). Spectral line cuts and conductance maps show no ring-like features (see Supplementary Figs. S8-S10). Consistently, magnetoresistance measurements reveal only a modest response (a few percent decrease) in {$EuCd_2Sb_2$}~\cite{Su2020} as compared to the pronounced ~90\% negative magnetoresistance observed in {$EuZn_2As_2$}~\cite{ZhiChenWang2022,Luo2023}.

To gain insight into the gap structure and topological properties of both compounds, we perform first-principles calculations using the Vienna Ab-initio Simulation Package (VASP)~\cite{Kresse1993,Kresse1996,KresseandFurth1996} to obtain their electronic band structures and corresponding partial density of states. To explore the importance of the electronic correlations, we compare results obtained from GGA+U, SCAN+U, and mBJ+U approaches, using interaction parameters of U=7eV, $J_H$=0.15U  introduced to 4f-electrons of Eu atoms in all the calculations. We mainly focus on the magnetically ordered state of AFM-A and include the spin-orbit coupling (see Supplementary Fig.S12 for the calculation details).

Figur 4 shows the band structure and surface dependent density of states using the mBJ+U approach. Results using the GGA+U and SCAN+U approaches can be found in the Supplementary Information. We find that while the GGA+U and SCAN+U methods fail to produce a band gap in {$EuCd_2Sb_2$} (in agreement with70), the mBJ+U approach successfully yields a band gap of approximately 120 meV (see Figure 4a), which is comparable to the result of ~200 meV obtained by the computationally expensive hybrid functional (HSE) approach70 and in reasonable agreement with experimentally observed gap. For {$EuZn_2As_2$}, we find that while the GGA+U approach still fails to produce a band gap, the SCAN+U and mBJ+U methods successfully yield gaps of approximately 350 meV and 890 meV, respectively (see Figure 4b and Supplementary Fig.S12). Notably, the gap obtained using mBJ+U (890 meV) is relatively close to the experimentally observed gap of ~1.2 eV. Therefore, mBJ+U method adequately predicts a reasonable band gap in both materials. Since the main effect of the Hubbard U interaction is to push the 4f levels away, the existence of the band gap in both materials is a result of the correlation effect of p and s orbitals on non-Eu atoms, which is exactly what mBJ method is designed to treat appropriately~\cite{Tran2009,Koller2012}.

The projected density of states (DOS) for each atom in both materials, obtained using the mBJ+U method, is shown in Figures 4a,4b and provides insight into the surface-dependent gaps observed in the experimental data. Since the bands near the Fermi energy are primarily composed of p and s orbitals from non-Eu atoms, the gap in the projected DOS on these atoms represents the actual band gap. In contrast, the gap observed in the projected DOS on Eu atoms corresponds to the energy separation between the 4f levels. For {$EuCd_2Sb_2$}, the true band gap is significantly smaller than the 4f level separation, leading to different gap values in the STM spectra for Eu- and Sb-terminated surfaces. In {$EuZn_2As_2$}, however, the real band gap is much larger and nearly matches the 4f level separation, resulting in a more consistent gap measurement across different surfaces.

Finally, we would like to address the band topology implied from our mBJ+U calculation. As pointed out in~\cite{Cuono2023}, the topology of the electronic structures can be indicated by a quantity defined at the $\Gamma$ point $E_g (\Gamma)=E_s^A (\Gamma)-E_p^X (\Gamma)$, where $E_s^A (\Gamma)$ is the conduction band minimum of s orbital of A ions in $EuA_2X_2$ (A=Cd,Zn), and $E_p^X (\Gamma)$ is the valence band maximum of p orbital of the X ions (X=Sb,As ). We find that both materials have $E_g (\Gamma)>0$, suggesting that no band inversion occurs in their band structures and consequently they are both topologically trivial. Therefore, our first principles calculations using mBJ+U method predicts no topological surface states in both materials, which is in an excellent agreement with the experimental data.

In summary, we have investigated the electronic properties of {$EuZn_2As_2$} and {$EuCd_2Sb_2$} through scanning tunneling spectroscopy, supported by first-principles calculations. Our results provide compelling evidence that both materials are trivial magnetic semiconductors, with no sign of topological band inversion. Notably, similar semiconducting characteristics in {$EuCd_2Sb_2$} was recently observed in optical spectroscopy~\cite{Nasrallah2024}. Finally, our topographic and spectroscopic data reveal charge localization at defect sites, clearly notable in {$EuZn_2As_2$}, which may promote the formation of magnetic polarons, offering a natural explanation for the large negative magnetoresistance observed in these materials~\cite{Goforth2008,ZhiChenWang2022,HCChen2020,Luo2023,Pervakov2024,YingLi2024,Balguri2024,Pohlit2018,Calderon2004,Alecrivillero2023,Souza2022,cook2025,kopp2025}.

\section*{Acknowledgments}
PA acknowledges funding from the U.S. National Science Foundation (NSF) under award No. DMR-2406686. DK and TR were supported by the National Science Centre (Poland) under research grant 2021/41/B/ST3/01141.

\newpage
\floatsetup[figure]{style=plain,subcapbesideposition=top}

\begin{figure}[h]
    \centering
        \includegraphics[width=1\linewidth]{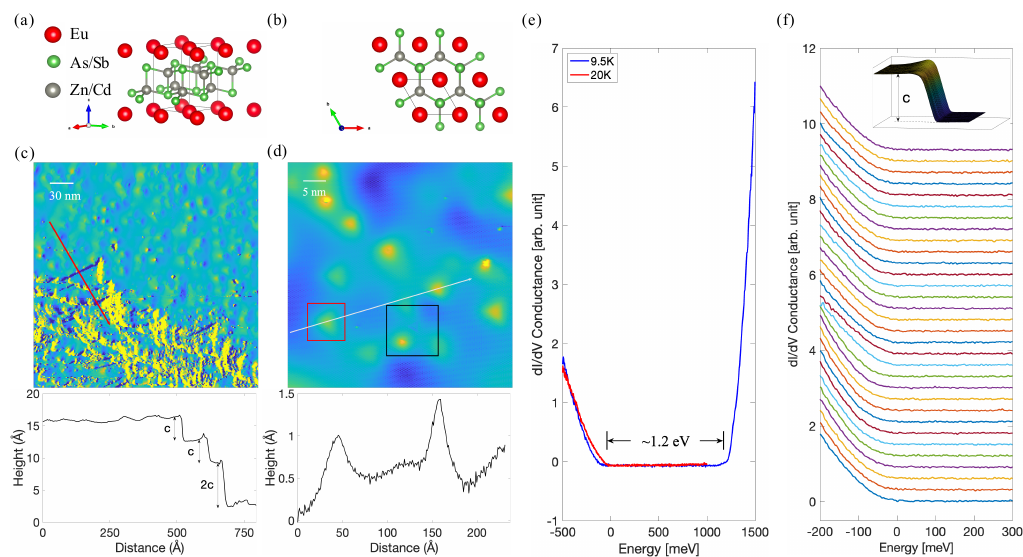}
    
    \caption{
     (a),(b) Crystal structure of {$EuZn_2As_2$} and {$EuCd_2Sb_2$}. 
     (c) Large scale STM topographic image of {$EuZn_2As_2$} sample. The lower inset shows a topographic linecut across multiple steps (indicated by red arrow) (set point: Bias : -100mV and Current = -400pA).
    (d) Smaller scale topographic image of {$EuZn_2As_2$} sample on the flat area. Lower inset shows a topographic linecut along the white arrow. Top inset shows the Fourier Transform of the topograph (set point: Bias : -200mV and Current = -50pA).
    (e) dI/dV conductance spectra taken on a flat area of {$EuZn_2As_2$} sample at 9.5K and 20K. 
    (f) dI/dV conductance spectra taken across a step edge (topography indicated in the inset) with step height of c.}
    \label{fig:figure1}
\end{figure}

\begin{figure}[h]
    \centering
        \includegraphics[width=1\linewidth]{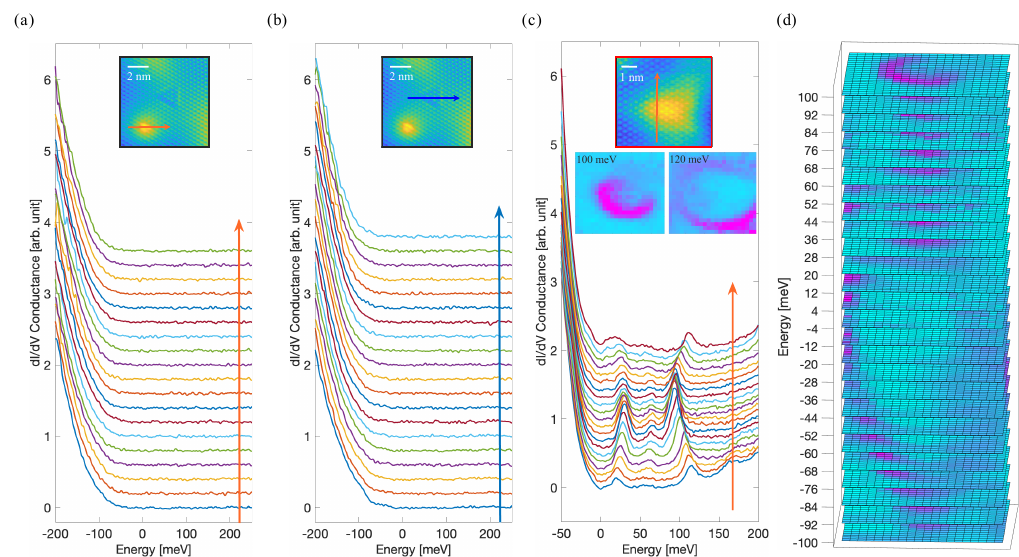}

     \caption{
     (a)-(c) dI/dV conductance spectra taken on the {$EuZn_2As_2$} sample along the arrow indicated by the topography shown in the insets (a)-(b) set point: Bias = -200mV and Current = -50pA, and (c) set point: Bias = -100mV and Current = -400pA), which are the areas indicated by black and red boxes in Figure (1d). Insets of (c) also shows the dI/dV conductance maps at +100 and +120 meV, which clearly reveals the ring-like structure with its radius varying at different energies. (d) Stacked dI/dV conductance maps taken at the area indicated in the top inset in (c) at various energies ranging from -100 to 100 meV, showing the evolution of the ring-like structure as a function of energy.}
     \label{fig:figure2}
\end{figure}

\begin{figure}[h]
    \centering
    \includegraphics[width=1\linewidth]{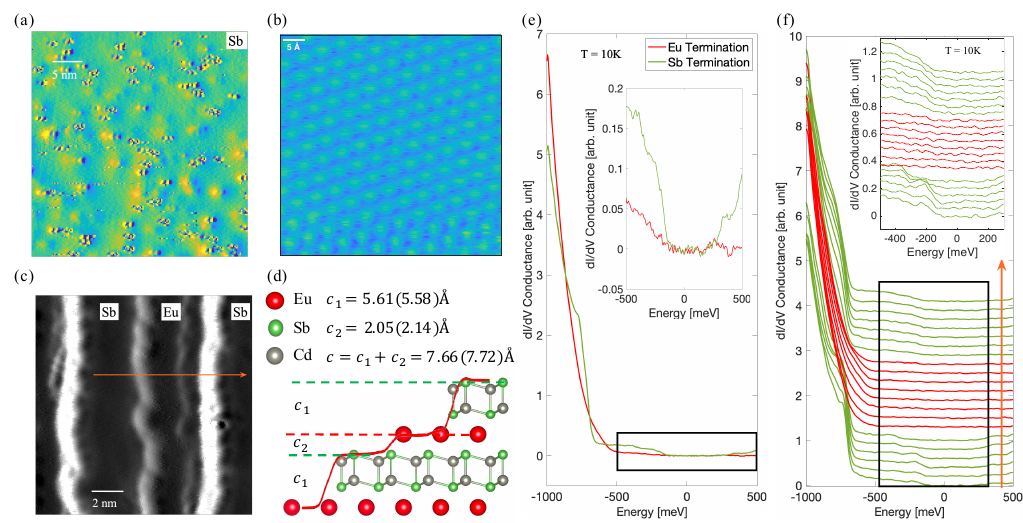}
    
    \caption{(a) Topographic image of {$EuCd_2Sb_2$} sample (set point: Bias = -1V and Current = -100pA). (b) Zoomed-in topograph on a flat terrace away from defects showing the atomic lattice (set point: Bias = -1V and Current = -100pA). (c) Topographic image of an area including sub-unit cell steps with two terminations. The horizontal line profile across the steps indicated by the red curve. (d) Side view of the crystal structure overlapped by the line profile in (c), with values of $c_1$ and $c_2$ measured by experiment (obtained from the crystal structure) indicated on top. (e) Comparison of dI/dV conductance spectra taken on the Sb- and Eu-terminated surfaces. The inset shows a zoom-in of the gap structure in the black box. (f) dI/dV conductance spectra taken along the arrow indicated in (c). Insets show the zoomed-in curves inside the black box.} 
     \label{fig:figure3}
\end{figure}

\begin{figure}[h]
    \centering
        \includegraphics[width=1\linewidth]{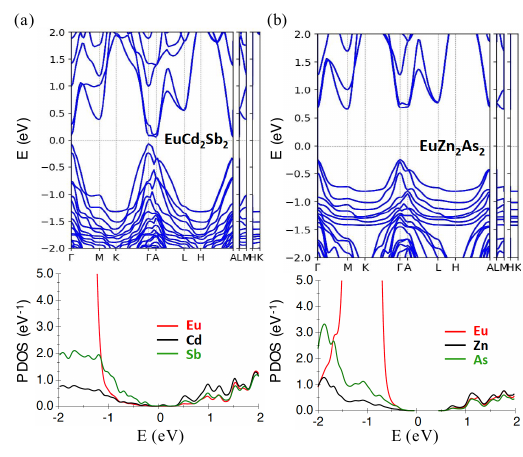}

 \caption{Electronic band structure and partial density of states calculated using the mBJ+U approach for (a) {$EuCd_2Sb_2$} and (b) {$EuZn_2As_2$}.}
     \label{fig:figure4}
\end{figure}
    
\clearpage
\bibliography{export}
\end{document}